\documentclass[conference]{IEEEtran}
\usepackage{url}
\usepackage{algorithm}
\usepackage{algorithmic}
\usepackage{ upgreek }
\ifCLASSINFOpdf
\usepackage[pdftex]{graphicx}
\else
\usepackage[dvips]{graphicx}
\fi
\usepackage{amsmath}
\usepackage{color}
\usepackage{ dsfont }
\usepackage{mathtools}

\usepackage[misc]{ifsym}
\usepackage{algorithmic}
\usepackage{array}
\usepackage{verbatim}
\ifCLASSOPTIONcompsoc
\usepackage[caption=false,font=normalsize,labelfont=sf,textfont=sf]{subfig}
\else
\usepackage[caption=false,font=footnotesize]{subfig}
\fi
\usepackage{amsmath}

\begin{document}

\title{Conductor Galloping Prediction on Imbalanced Datasets: SVM with Smart Sampling}

\author{\IEEEauthorblockN{Kui Wang, Jian Sun, Chenye Wu, \textit{and} Yang Yu$^\text{\Letter}$}
\IEEEauthorblockA{Institute for Interdisciplinary Information Sciences\\ 
Tsinghua University, Beijing, 100084, P.R. China\\
Email: yangyu1@tsinghua.edu.cn
}
}
\maketitle
\begin{abstract}
Conductor galloping is the high-amplitude, low-frequency oscillation of overhead power lines due to wind. Such movements may lead to severe damages to transmission lines, and hence pose significant risks to the power system operation. In this paper, we target to design a prediction framework for conductor galloping. The difficulty comes from imbalanced dataset as galloping happens rarely. By examining the impacts of data balance and data volume on the prediction performance, we propose to employ proper sample adjustment methods to achieve better performance. Numerical study suggests that using only three features, together with over sampling, the SVM based prediction framework achieves an F$_1$-score of 98.9$\%$\footnote{This work has been supported in part by National Key R\&D Program of China (2018YFC0809400), the Youth Program of National Natural Science Foundation of China (No. 71804087), and Turing AI Institute of Nanjing.}.
% Conductor galloping may incur serious damages to transmission lines and usually happens in extreme weather. However, developing an efficient prediction model for conductor galloping remains as a challenging problem due to low-quality historical data collecting and incomplete analysis. In this paper we first check weather features which really related to conductor galloping, then analyze how the balance level and volume of the data influence the prediction performance. At last, with data re-sampling methods, we give our efficient prediction model with a 98.9\% F$_1$-score by using a moderate training dataset with 63000 samples and only 3 features.
\end{abstract}
\vspace{0.1cm}


\IEEEpeerreviewmaketitle
\section{Introduction}

\begin{figure}[t]
    \centering
    \includegraphics[width=3in]{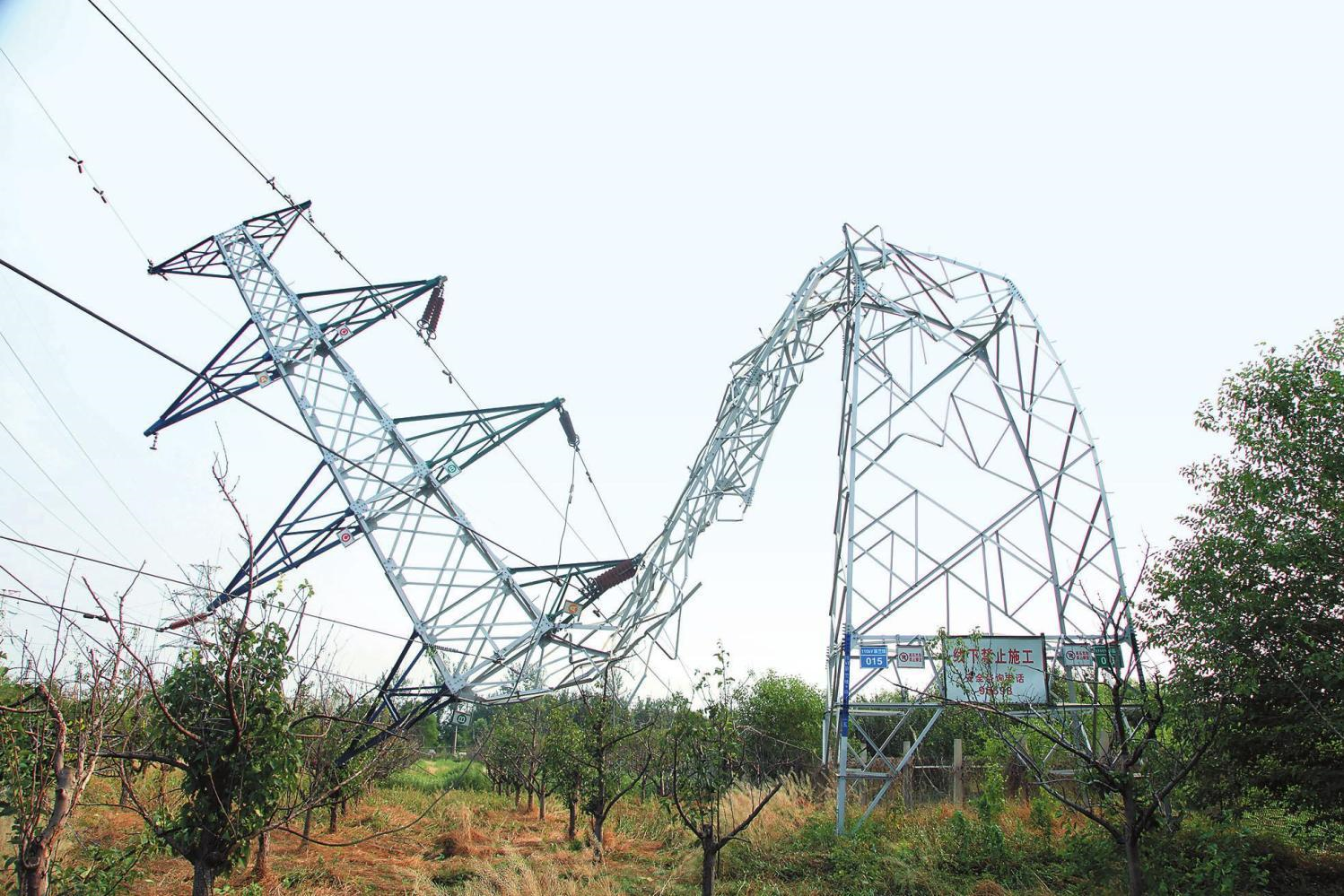}
    \caption{Pole collapse induced by conductor galloping \cite{Pole}.}\vspace{-0.5cm}
    \label{fig:damage}
\end{figure}

Most modern power system control frameworks are designed for (near) normal operation conditions, which makes power system vulnerable to the risks posed by extreme weathers. Hence, the damage prediction due to extreme weather is critical to maintain a reliable power system. In this paper, we target to design the prediction framework for conductor galloping, which often happen in windy and humid conditions, and may result in huge damages to the power grid, such as tripping, bolt looseness and even pole collapse accident \cite{Zhu&gallopingdamage}. Such damages do happen. In June, 2018, extreme hail wind caused conductor galloping, which resulted in severe pole collapse accident (as shown in Fig. \ref{fig:damage}) in Shunyi District, Beijing, P.R. China.

The conventional wisdom for conductor galloping prediction is to use a model based approach, which examines the physical process of galloping and analyzes the trigger conditions in the process. The seminal work is the transmission line vibration model, proposed by Den Hartog in 1932 \cite{den1932transmission}. Based on this model, Nigol and Clarke introduced the physical process of galloping and set the stage in 1974 \cite{nigol1974conductor}. Since then, there are only minor modifications towards the model based understanding of galloping. With the advance in sensing technology, it is now possible to take a data-driven approach for galloping prediction.

\subsection{Challenges and Opportunities}
Forecasting extremal event is a challenge for the data science, primarily due to extremely small sample size. For instance, it is hard to improve the predication accuracy for the critical peak load, while the general load forecast has already been accurate \cite{loadprediction}. In the most recent study, the F$_1$-score of galloping prediction is far from satisfactory due to the limited data availability (e.g. 83\% in \cite{Cheng&smalldata}). Such limitation also challenges a wide range of emerging algorithmic technologies, such as machine translation \cite{Machinetranslation} and recommendation systems \cite{googlepagerank}. Thus, improving the extremal event forecast based on limited data is widely beneficial.

The most recent deployment of the smart grid monitoring meters across North China has collected a sizable dataset for conductor galloping. In this research, we seek to develop a data-driven prediction model based on this dataset. Specifically, we adopt models to facilitate the investigation on the tensions between three questions:
\begin{enumerate}
    \item Which features are important in the prediction model?
    \item How will the data imbalance affect the prediction accuracy under different data volume?
    \item What is the role of dataset volume in galloping prediction?
\end{enumerate}

\begin{figure}[t]
    \centering
    \includegraphics[width=3.2in]{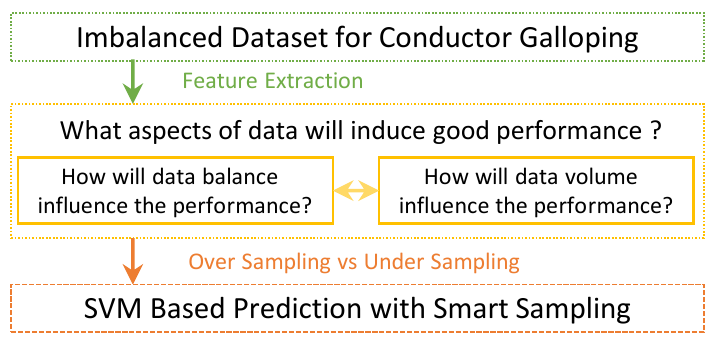}
    \caption{Our framework for galloping prediction analysis.}
    \label{fig:diagram}
\end{figure}

The answers inspire us to propose the smart sampling approach, which improves the dataset quality, yielding a better prediction accuracy. Figure \ref{fig:diagram} plots the paradigm of our efforts towards designing the prediction framework.

\subsection{Literature Review}
Towards answering the aforementioned three questions, we identify two major related research directions. The first one focuses on feature extraction, which contains a rather rich literature. We refer interested readers to an excellent survey \cite{chandrashekar2014survey} for more details.

Another related research direction investigates the data imbalance issues in machine learning. For example, Liu \emph{et al.} conduct an empirical study to highlight how class-imbalance affects the performance of cost-sensitive classifiers in \cite{LiuX}. Take support vector machine (SVM) as an example, Much efforts have been spent towards tackling the challenges in applying SVM on imbalanced datasets: utilizing the information-loss-minimization principle \cite{Tang}, adjusting the class boundary based on kernel-boundary alignment algorithm \cite{Wu}, \emph{etc}.

The research on designing customized machine learning algorithm for conductor galloping prediction is very limited. To the best of our knowledge, we are the first to design the prediction framework with an emphasis on understanding how to best utilize the information in the imbalanced dataset. 

\subsection{Our Contributions}
In seek of designing the customized prediction framework for conductor galloping, our principal contributions can be summarized as follows:
\begin{itemize}
    \item \emph{Feature Extraction}: We identify the determinants to trigger conductor galloping by observing the data distribution and validate our observations with the model based approach. The prediction model with the identified features achieves an F$_1$-score of 98.9$\%$.
    \item \emph{Assess the Value of Data}: By proposing the prediction framework, we investigate how the imbalance in the dataset limits the prediction performance, which in turn reveals the true value of heterogeneity in a dataset.
    \item \emph{Smart Sampling Approach:} We design a smart sampling approach to improving the  dataset quality for better prediction accuracy. We highlight via numerical studies the value of this approach when the volume of the dataset is limited.
\end{itemize}

The rest of our paper is organized as follows. Section \ref{sec:data} overviews our galloping dataset and revisits the theoretical models for galloping detection. In this paper, we choose SVM for the prediction framework and we introduce the evaluation metrics and feature extraction in Section \ref{sec:algo}. From extensive numerical studies in Section \ref{sec:result}, we exploit the sample adjustment approach for imbalanced datasets to achieve better prediction accuracy. Finally, concluding remarks and future directions are given in Section \ref{sec:con}.

\section{Data and Model: the Basics}
\label{sec:data}

\begin{figure}[t]
    \centering
    \includegraphics[width=3.4in]{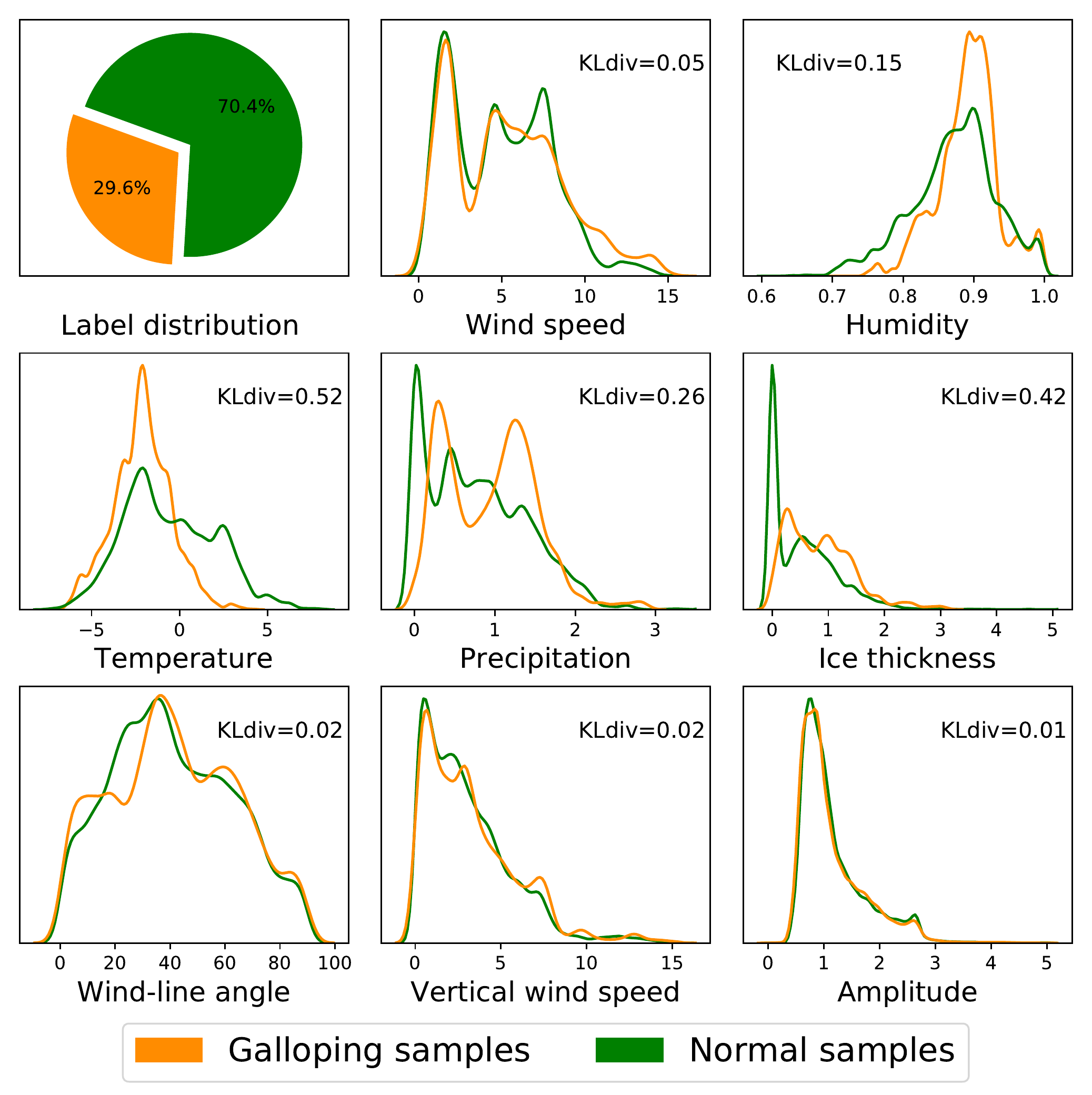}
    \caption{The distributions for features in galloping and normal samples. The ``KLdiv" refers to the KL divergence of feature distribution in the two sample groups.}
    \label{fig:dist}
\end{figure}

The historical galloping data is collected from October 2017 to January 2018 in China. Among the 80,596 meter collected samples, 25,414 pieces are galloping samples. We plot the distributions of 8 features (wind speed, humidity, temperature, precipitation, ice-thickness, wind-line angle, vertical wind speed and amplitude) in Fig. \ref{fig:dist}. This figure indicates that besides imbalanced sample size (only 30\% galloping data), the feature distributions show even more severe imbalance. This highlights the urgent need for customizing the prediction framework for galloping. Note that for the long distance overhead power lines, the wind-line angle is generally not very well defined. Hence, in the subsequent analysis, we only use the other 7 features.
%It's worth noting that not only the sample size are different between the galloping (about 30\%) and normal samples (about 70\%), but also the features' distribution. For example, the distribution of precipitation in normal samples has a much more sharply peak around zero, which intuitively means that less precipitation can suppress conductor galloping.
%We can get some inspirations from the theoretical models for exploring which features will really influence the conductor galloping. Den Hartog \cite{den1932transmission} firstly introduce the theoretical analysis to conductor galloping. In his model, galloping happens if the iced conductor face a proper wind condition which can be represented as:

\begin{figure*}[t]
\centering
    \includegraphics[width=6.4in]{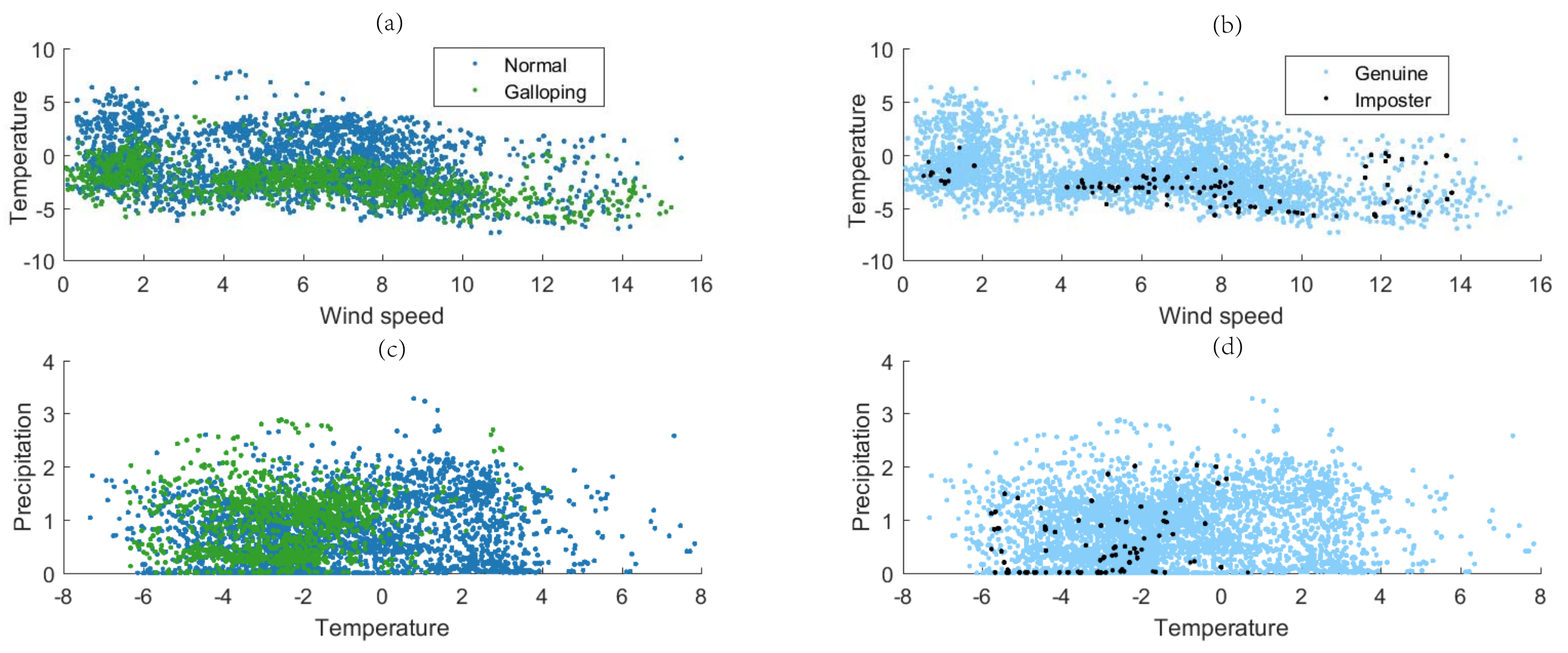}
    \caption{Distributions of the two sample groups in 2-feature hyper planes. (a) and (c) demonstrate the distribution of samples' true labels in the wind speed-temperature plane and temperature-precipitation plane, respectively. (b) and (d) highlight the imposters in this prediction model.}
    \label{fig:boundary}
\end{figure*}

Before directly diving into the machine learning analysis, we would like to first revisit the theoretical model for conductor galloping \cite{den1932transmission}: the trigger condition is as follows:
\begin{align}
    \frac{\partial C_L}{\partial \alpha}+C_D<0,
\end{align}
where $C_L$ and $C_D$ are the aerodynamic lift and drag coefficient of the conductor, and $\alpha$ is the angle of attack from wind. 
Note that $C_L$ and $C_D$ are also functions of wind speed and wind line angle. This theoretical result gives us the first cut in identifying the important features for the prediction framework.

\section{Metrics and Feature Extractions}
\label{sec:algo}
In this paper, we select a conventional SVM model with the Gaussian kernel \cite{Keerthi} to establish the prediction framework.
\subsection{Performance Metrics}
In the machine learning literature, the most widely adopted metrics for performance evaluation are \emph{recall} and \emph{precision}. \emph{Recall} measures that among the galloping samples, how well our predictor can predict the extremal events. Denote the true label and predicted label of the predictor for sample $i$ by $y_i$ and $\hat{y}_i$, respectively. And assume galloping samples are labelled by 1 while the other samples are labelled by -1. Then,
\begin{equation}
    \text{Recall}=\frac{\sum_i I(\hat{y}_i=1)\times I(y_i=1) }{\sum_i I(y_i=1)},
\end{equation}
where $I(\cdot)$ is the indicator function.

On the other hand, \emph{precision} measures that among those samples being predicted to gallop, how many of them are actually with true label of galloping. More precisely,
\begin{equation}
    \text{Precision}=\frac{\sum_i I(\hat{y}_i=1)\times I(y_i=1) }{\sum_i I(\hat{y}_i=1)}.
\end{equation}
In this paper, we adopt a single metric F$_1$-score to measure the comprehensive performance over \emph{recall} and \emph{precision} \cite{Yang:1999:RTC:312624.312647}, which is defined as follows:
\begin{equation}
    \text{F}_1\text{-score}=\frac{2\times \text{Precision}\times \text{Recall}}{\text{Precision}+\text{Recall}}.
\end{equation}
The test set is randomly selected from the whole dataset (at a 25\% proportion), disjoint from the training set.

\subsection{Feature Extraction}
\begin{figure}[t]
    \centering
    \includegraphics[width=3.2in]{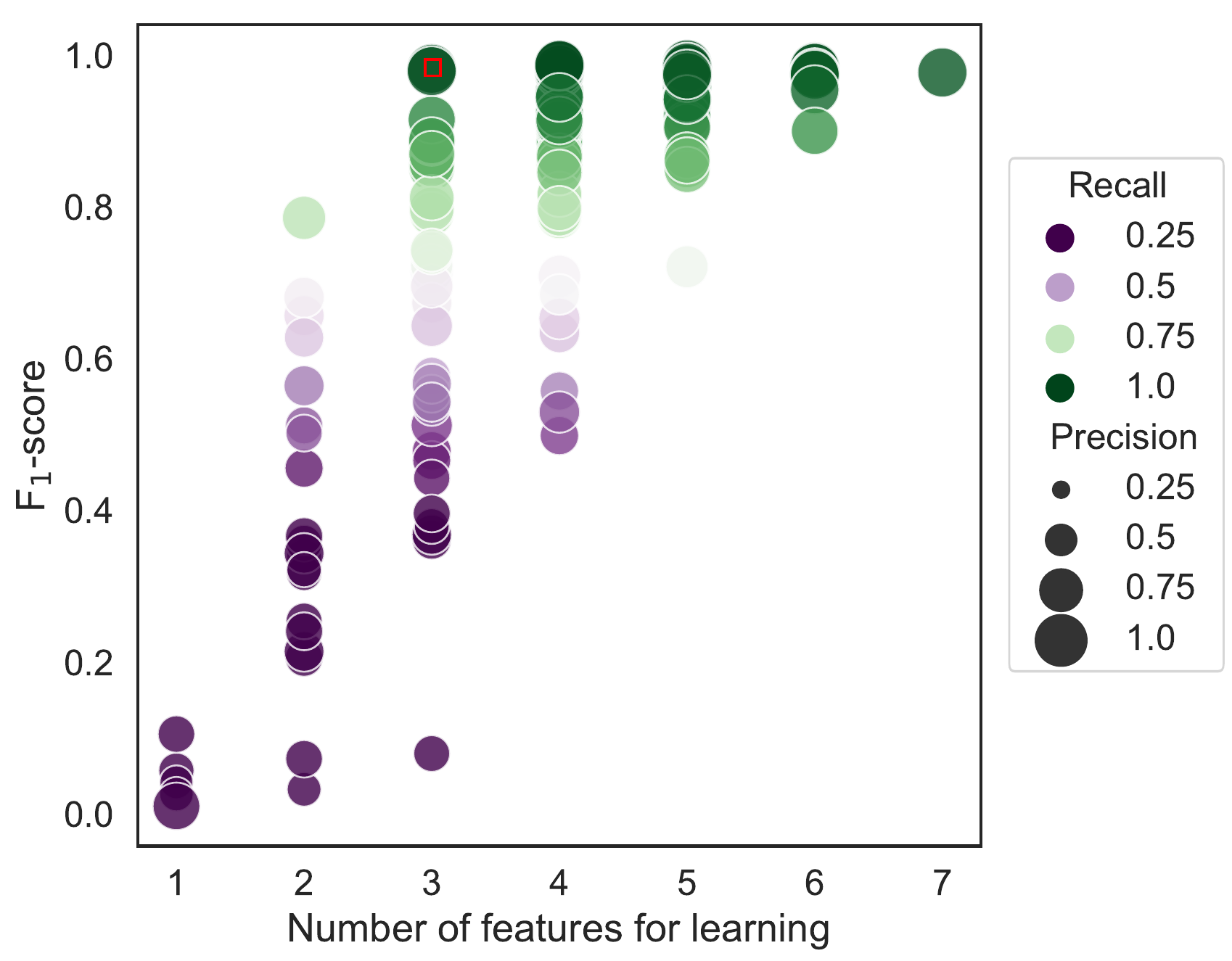}
    \caption{The prediction performance of different feature groups.}
    \label{fig:featureimportance}
\end{figure}

To extract the most powerful set of features on the conductor-galloping prediction, we test the F$_1$-score performance of all the 127 possible combinations among the 7 features. Surprisingly, we can use only three features and achieve remarkably good performance with an F$_1$-score of 98.4$\%$. These three features are wind power, temperature and precipitation. Figure \ref{fig:boundary} plots the projections of the samples onto two hyper planes. It is evident that with respect to wind speed and precipitation, temperature is a good classifier. Figure \ref{fig:boundary} also shows the classification results with minor imposters. We illustrate the performance of all the 127 combinations in Fig. \ref{fig:featureimportance}. While more features improve the F$_1$-score, the improvement, compared with the selected three features, is only marginal.

\begin{figure}[t]
    \centering
    \includegraphics[width=3.2in]{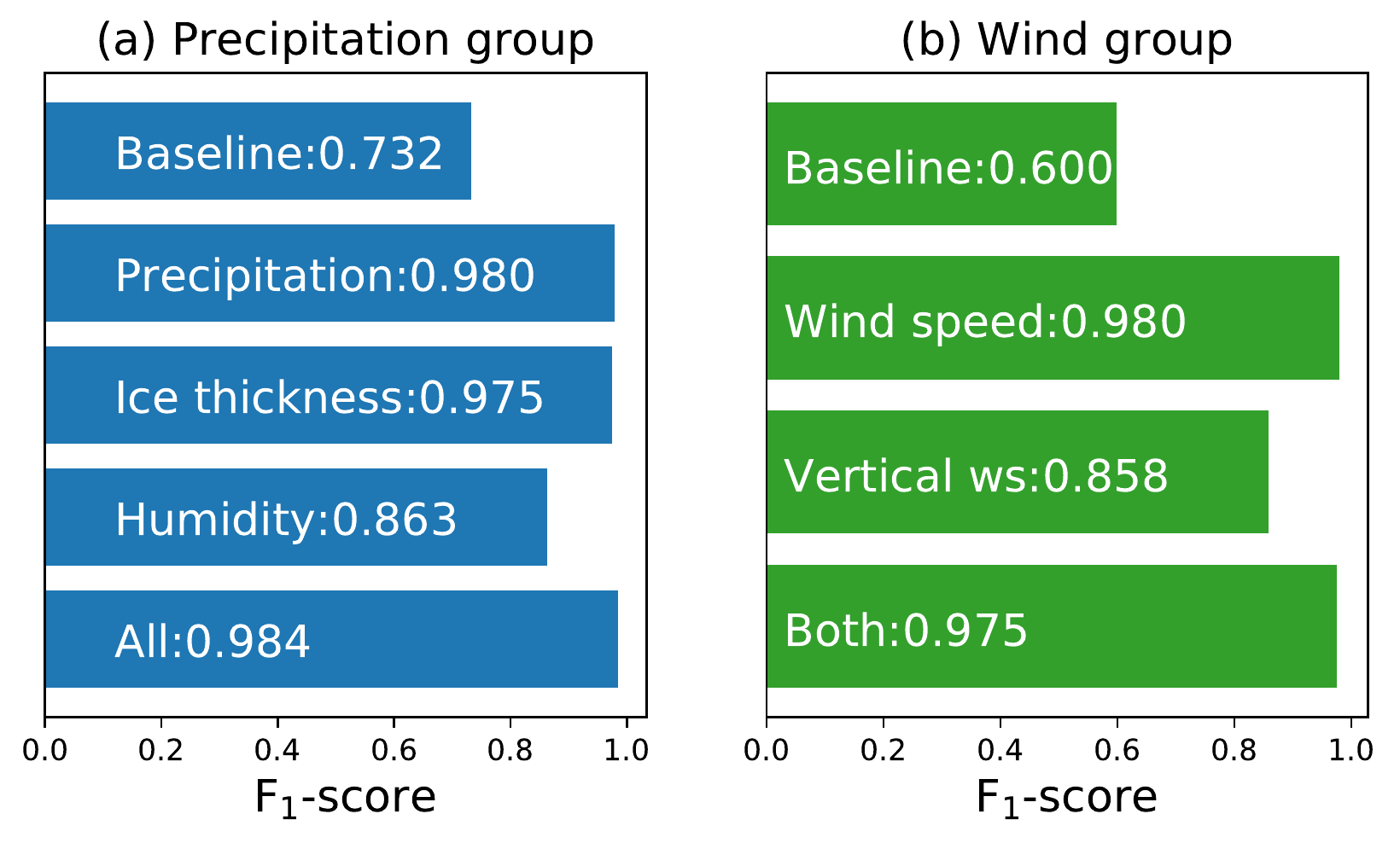}
    \caption{The substitute relationship among features.} 
    \label{fig:twogroupshow}
\end{figure}

To further exploit the substitute structure in the 7 features, we start our analysis on the selected three features set. We first seek to understand the possible substitute for precipitation. With only two features, wind speed and temperature, the SVM model can achieve an F$_1$-score of 73.2$\%$. Including any one of precipitation, ice-thickness and humidity will increase the F$_1$-score dramatically. In particular, from the final performance, precipitation and ice-thickness are prefect substitute features for conductor galloping prediction. This also aligns with our intuition. Since most conductor galloping events happen at the temperature around 0$^\circ$C, in this condition, precipitation and ice-thickness on the overhead power line are closely correlated. In this regard, humidity is also a good substitute of precipitation and ice-thickness. However, including all the three features won't further improve the F$_1$-score dramatically. Figure \ref{fig:twogroupshow} (a) visualizes this substitute relationship. We can conduct the same analysis to understand the substitute relationship between wind speed and vertical wind speed. We visualize the result in Fig. \ref{fig:twogroupshow} (b). One interesting observation is that vertical wind speed is believed to be a more important determinant in triggering galloping as suggested by Den Hartog's model \cite{den1932transmission}. However, in practice, for a long-distance overhead power line, the vertical wind speed is also not very well defined, which can help us to explain why vertical wind speed plays a weaker role in the SVM prediction model compared with wind speed.

\section{Sampling for Imbalanced Datasets}
\label{sec:result}
A rich literature has suggested that the performance of a forecast model trained by limited data is contingent on whether the samples are balanced over features as well as whether they can represent the population's distribution. This inspires us to conduct sample adjustment to balance the dataset over features for better performance.

\begin{figure}[t]
    \centering
    \includegraphics[width=3.2in]{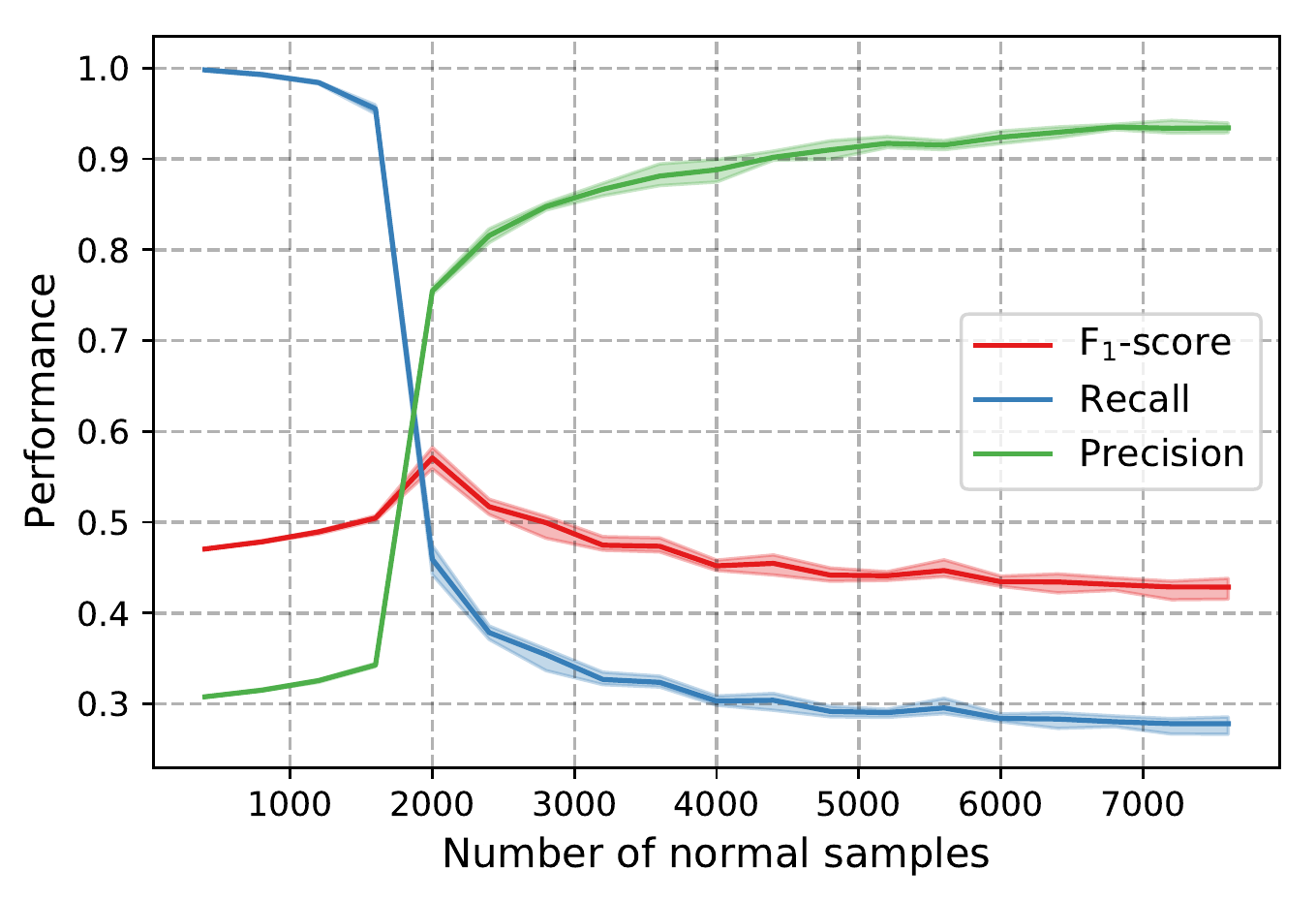}
    \caption{The prediction performance varies according to number of normal samples in the training set.}
    \label{fig:balance}
\end{figure}

\begin{figure}[t]
    \centering
    \includegraphics[width=2.8in]{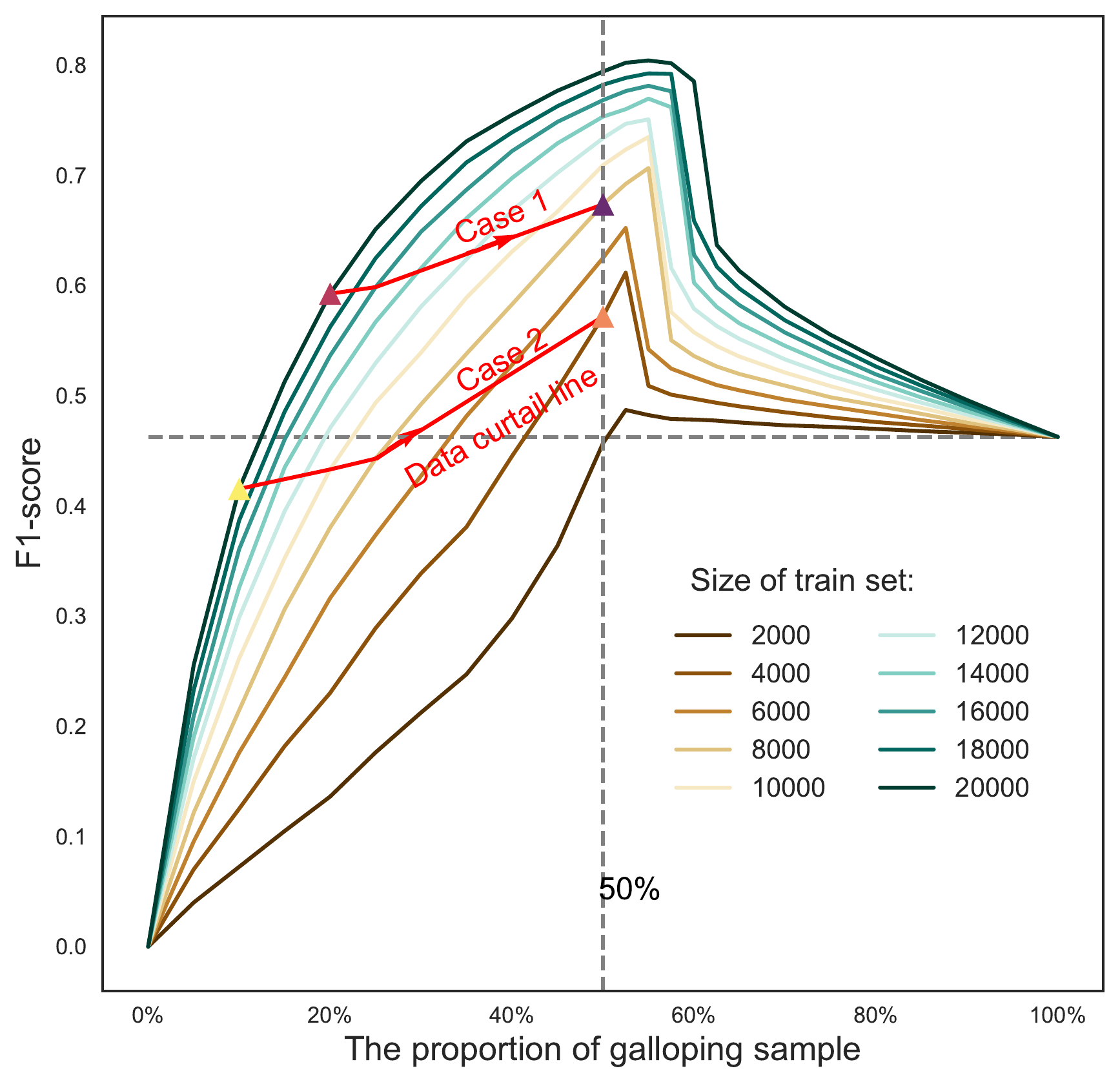}
    \caption{The impact of data imbalance on the prediction performance.}
    \label{fig:balanceandsize}
\end{figure}

\begin{figure*}[!h]
    \centering
    \includegraphics[width=7in]{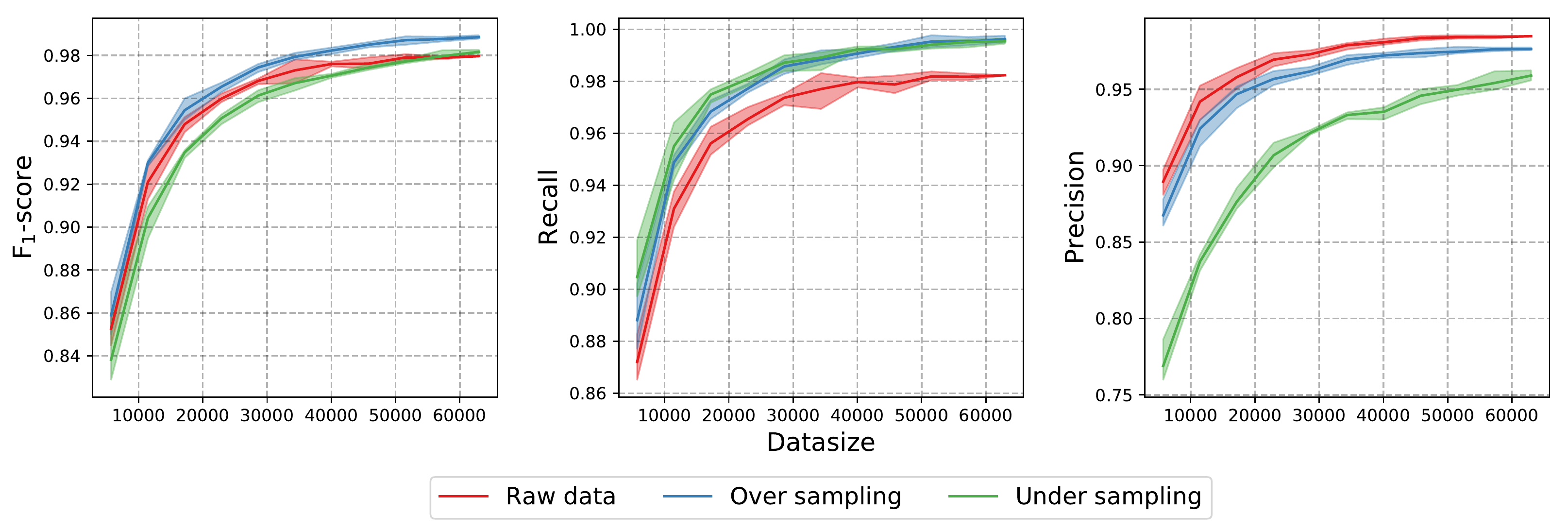}
    \caption{Performance comparison with different data balancing methods. (a) F$_1$-score; (b) Recall; (c) Precision.}
    \label{fig:bestmodel}
\end{figure*}

\subsection{Role of Data Balance}
We first investigate the role of data balance in galloping prediction: we examine the prediction performance by constructing a dataset including 2,000 galloping samples and an increasing number of normal samples. Figure \ref{fig:balance} plots the evolving performance. In this case, more normal samples in the training data will increase the precision while decrease the recall. The best trade-off illustrated by F$_1$-score happens in the data balance point (same amount of galloping and normal samples).

It is interesting to note that too much normal samples in the training set even decreases the performance. We conduct more numerical studies to highlight this observation: for the size of the dataset ranging from 2,000 to 20,000, Fig. \ref{fig:balanceandsize} investigates the value of data balance for better performance. The peak performance is achieved almost always at the data balance point. Based on Fig. \ref{fig:balanceandsize}, we make a few more observations on the value of dataset volume: given the same imbalanced level, larger volume implies better performance. On the other hand, larger volume dataset is also more robust to data imbalance.
\subsection{Sampling for Data Balance}
We focus on the selected set of three features: wind speed, temperature and precipitation. To achieve a better performance, we employ two sample adjustment methods to balance the dataset: under sampling and over sampling.

For under sampling, we drop some normal samples in the 
dataset to achieve the balance while simultaneously maintaining the distribution of each feature in the normal sample group unchanged. To achieve this goal, we just randomly select a normal sample in current dataset by index and drop it, and repeat this process until the dataset is balanced.

For over sampling, we use the Synthetic Minority Over-sampling Technique (SMOTE) \cite{chawla2002smote} methods to boost the galloping samples. SMOTE is frequently used for balancing imbalanced dataset for machine learning. To over sample the galloping samples (the minority) in the dataset by SMOTE, we first take a real galloping sample from it, and find its $k$ nearest galloping neighbors in the feature space. Then we randomly select one of its neighbors and create a sample by a random linear combination of the selected galloping sample and its galloping neighbor. Finally we add this new point with the label of galloping to the current dataset. By repeating this process, we can create many new galloping samples to balance the dataset.

Figure \ref{fig:bestmodel} compares the performance of the two methods. Both methods increase recall and decrease precision. For galloping prediction task, over-sampling outperforms the other two rivals, and achieves an F$_1$-score of 98.9$\%$. This is because in the original dataset, though imbalanced, the galloping samples still accurately represent the true distribution of the galloping population, which allows us to conduct valid over-sampling.

\section{Conclusion}
\label{sec:con}
In this paper, we design the SVM prediction framework for conductor galloping prediction. To improve the prediction performance, we examine the impacts of data balance and data volume on F$_1$-score, and we submit that a balanced dataset is vital to achieve a remarkably good performance with limited resources. For the purpose of conductor galloping prediction, numerical studies suggest that over sampling is a good approach to maintaining the data balance.

This work can be extended in many ways. For example, it is important to examine the ability of generalization of our proposed model, as it is generally costly to collect data for extremal events. We also intend to design an adaptive online learning framework for conductor galloping prediction.
\bibliographystyle{IEEEtran}
\bibliography{IEEEabrv,ref}

\begin{thebibliography}{10}
\providecommand{\url}[1]{#1}
\csname url@samestyle\endcsname
\providecommand{\newblock}{\relax}
\providecommand{\bibinfo}[2]{#2}
\providecommand{\BIBentrySTDinterwordspacing}{\spaceskip=0pt\relax}
\providecommand{\BIBentryALTinterwordstretchfactor}{4}
\providecommand{\BIBentryALTinterwordspacing}{\spaceskip=\fontdimen2\font plus
\BIBentryALTinterwordstretchfactor\fontdimen3\font minus
  \fontdimen4\font\relax}
\providecommand{\BIBforeignlanguage}[2]{{%
\expandafter\ifx\csname l@#1\endcsname\relax
\typeout{** WARNING: IEEEtran.bst: No hyphenation pattern has been}%
\typeout{** loaded for the language `#1'. Using the pattern for}%
\typeout{** the default language instead.}%
\else
\language=\csname l@#1\endcsname
\fi
#2}}
\providecommand{\BIBdecl}{\relax}
\BIBdecl

\bibitem{Zhu&gallopingdamage}
{Kuan-jun Zhu}, {Bin Liu}, {Hai-jun Niu}, and {Jun-hui Li}, ``Statistical
  analysis and research on galloping characteristics and damage for iced
  conductors of transmission lines in china,'' in \emph{2010 International
  Conference on Power System Technology}, Oct 2010, pp. 1--5.

\bibitem{den1932transmission}
J.~Den~Hartog, ``Transmission line vibration due to sleet,'' \emph{Transactions
  of the American Institute of Electrical Engineers}, vol.~51, no.~4, pp.
  1074--1076, 1932.

\bibitem{nigol1974conductor}
O.~Nigol and G.~Clarke, ``Conductor galloping and control based on torsional
  mechanism,'' in \emph{IEEE Transactions on Power Apparatus and Systems},
  no.~6, 1974, pp. 1729--1729.

\bibitem{Pole}
\BIBentryALTinterwordspacing
{China Electric Power News}. (2018) Pole collapse induced by conductor
  galloping. [Online]. Available: \url{http://alturl.com/cp7hw}
\BIBentrySTDinterwordspacing

\bibitem{loadprediction}
A.~Sinha and J.~Mondal, ``Dynamic state estimator using ann based bus load
  prediction,'' \emph{IEEE Transactions on Power Systems}, vol.~14, no.~4, pp.
  1219--1225, 1999.

\bibitem{Cheng&smalldata}
Y.~{Cheng}, J.~{Han}, J.~{Zhang}, and W.~{Hao}, ``Reconstructing the problem of
  galloping monitoring of traditional complex analytical mechanism into a
  prediction method for machine learning algorithm modeling,'' in \emph{2018
  IEEE 3rd Advanced Information Technology, Electronic and Automation Control
  Conference (IAEAC)}, Oct 2018, pp. 2552--2557.

\bibitem{Machinetranslation}
K.~{Chen}, T.~{Zhao}, M.~{Yang}, L.~{Liu}, A.~{Tamura}, R.~{Wang},
  M.~{Utiyama}, and E.~{Sumita}, ``A neural approach to source dependence based
  context model for statistical machine translation,'' \emph{IEEE/ACM
  Transactions on Audio, Speech, and Language Processing}, vol.~26, no.~2, pp.
  266--280, Feb 2018.

\bibitem{googlepagerank}
S.~Brin and L.~Page, ``The anatomy of a large-scale hypertextual web search
  engine,'' \emph{Computer networks and ISDN systems}, vol.~30, no. 1-7, pp.
  107--117, 1998.

\bibitem{chandrashekar2014survey}
G.~Chandrashekar and F.~Sahin, ``A survey on feature selection methods,''
  \emph{Computers \& Electrical Engineering}, vol.~40, no.~1, pp. 16--28, 2014.

\bibitem{LiuX}
X.~{Liu} and Z.~{Zhou}, ``The influence of class imbalance on cost-sensitive
  learning: An empirical study,'' in \emph{Sixth International Conference on
  Data Mining (ICDM'06)}, Dec 2006, pp. 970--974.

\bibitem{Tang}
Y.~{Tang}, Y.~{Zhang}, N.~V. {Chawla}, and S.~{Krasser}, ``Svms modeling for
  highly imbalanced classification,'' \emph{IEEE Transactions on Systems, Man,
  and Cybernetics, Part B (Cybernetics)}, vol.~39, no.~1, pp. 281--288, Feb
  2009.

\bibitem{Wu}
G.~{Wu} and E.~Y. {Chang}, ``Kba: kernel boundary alignment considering
  imbalanced data distribution,'' \emph{IEEE Transactions on Knowledge and Data
  Engineering}, vol.~17, no.~6, pp. 786--795, June 2005.

\bibitem{Keerthi}
S.~Keerthi and C.-J. Lin, ``Asymptotic behaviors of support vector machines
  with gaussian kernel,'' 2003.

\bibitem{Yang:1999:RTC:312624.312647}
\BIBentryALTinterwordspacing
Y.~Yang and X.~Liu, ``A re-examination of text categorization methods,'' in
  \emph{Proceedings of the 22Nd Annual International ACM SIGIR Conference on
  Research and Development in Information Retrieval}, ser. SIGIR '99.\hskip 1em
  plus 0.5em minus 0.4em\relax New York, NY, USA: ACM, 1999, pp. 42--49.
  [Online]. Available: \url{http://doi.acm.org/10.1145/312624.312647}
\BIBentrySTDinterwordspacing

\bibitem{chawla2002smote}
N.~V. Chawla, K.~W. Bowyer, L.~O. Hall, and W.~P. Kegelmeyer, ``Smote:
  synthetic minority over-sampling technique,'' \emph{Journal of artificial
  intelligence research}, vol.~16, pp. 321--357, 2002.

\end{thebibliography}

\end{document}